# Formation process and superparamagnetic properties of (Mn,Ga)As nanocrystals in GaAs fabricated by annealing of (Ga,Mn)As layers with low Mn content

Janusz Sadowski[1,2*], Jaroslaw Z. Domagala[2], Roland Mathieu[3], András Kovács[4], Takeshi Kasama[4], Rafal E. Dunin-Borkowski[5,4] and Tomasz Dietl[2,6]

[1] MAX-Lab, Lund University, P.O. Box 118, SE-221 00 Lund, Sweden

[2] Institute of Physics, Polish Academy of Sciences, al. Lotników 32/46, PL-02-668 Warszawa, Poland

[3] Department of Engineering Sciences, Uppsala University, P.O. Box 534, SE-751 21 Uppsala, Sweden

[4] Center for Electron Nanoscopy, Technical University of Denmark, DK-2800 Kongens Lyngby, Denmark

[5] Ernst Ruska-Centre and Peter Grünberg Institute, Research Centre Jülich, 52425 Jülich, Germany

[6] Faculty of Physics, University of Warsaw, PL-00-681 Warszawa, Poland

**Abstract**

[*] corresponding author    e-mail:  janusz.sadowski@maxlab.lu.se

X-ray diffraction, transmission electron microscopy, and magnetization measurements are employed to study the structural and magnetic properties of Mn-rich (Mn,Ga)As nanocrystals embedded in GaAs. These nanocomposites are obtained by moderate (400°C) and high temperature (560 and 630°C) annealing of (Ga,Mn)As layers with Mn concentrations between 0.1 and 2%, grown by molecular beam epitaxy at 270°C. Decomposition of (Ga,Mn)As is already observed at the lowest annealing temperature of 400°C for layers with initial Mn content of 1% and 2%. Both cubic and hexagonal (Mn,Ga)As nanocrystals, with similar diameters of 7 - 10 nm are observed to coexist in layers with an initial Mn content of 0.5% after higher temperature annealing. Measurements of magnetization relaxation in the time span 0.1 – 10 000 s provide evidence for superparamagnetic properties of the (Mn,Ga)As nanocrystals, as well as for the absence of spin-glass dynamics. These findings point to weak coupling between nanocrystals even in layers with the highest nanocrystal density.

## 1. INTRODUCTION

Prospects for fabricating application-ready spintronic devices by using dilute magnetic semiconductors (DMSs) are facing difficulties due to the lack of a suitable material that maintains ferromagnetic properties up to room temperature.[1,2] As a result there is renewed interest in nanocomposite systems consisting of ferromagnetic nanocrystals embedded in a semiconductor matrix,[2-6] for which a number of functionalities has been predicted[4,5] and observed.[6] Ferromagnetic nanocrystals can be formed by crystallographic phase separation, i.e. by the precipitation of a transition metal (TM) compound (to form a condensed magnetic semiconductor) or a TM metal.[2,3] Interestingly, in many cases chemical phase separation (often referred to as spinodal decomposition[4]) occurs, to form nanoscale regions that are rich in TM cations without any difference in crystallographic structure from the surrounding semiconductor matrix, making them difficult to detect.[2-5] The efficiency of the decomposition may depend on the position of the Fermi level in the semiconducting host.[7,8] The occurrence of a decomposition can also be difficult to identify. A combination of several experimental techniques is necessary to address this question.

In (Ga,Mn)As – one of the most comprehensively studied DMS materials with hole induced ferromagnetism,[1] phase separation results in the formation of Mn-rich nanocrystals that have high temperature ferromagnetic properties (with critical temperatures in the range 300 – 350 K). In this system, phase separation can be achieved in a controlled manner by the high temperature (HT) annealing of (Ga,Mn)As ternary alloy layers grown by molecular beam epitaxy (MBE) at low temperature (LT). LT MBE growth of chemically uniform (Ga,Mn)As layers takes place at substrate temperatures in the range 170 – 300$^o$C. HT post-growth annealing, leading to detectable phase separation is then performed at temperatures of 400 – 700$^o$C. After HT annealing, Mn-rich nanocrystals inside the GaAs matrix can be identified by transmission electron microscopy (TEM) techniques. The physical parameters of the nanocrystals, such as their sizes and densities, can be controlled by the annealing temperature used and by the initial Mn content. Thus HT annealed (Ga,Mn)As can be used as a model system for studying phase separation in TM doped DMS materials.

In this paper, we study the properties of (Mn,Ga)As:GaAs phase-separated material produced by moderate-to-high temperature annealing of LT MBE-grown (Ga,Mn)As ternary alloys. Although many literature reports on this system have been published, both prior to[12] and after[13-15] the first successful growth of (Ga,Mn)As ferromagnetic semiconductor,[18] there are still unexplained issues concerning both the properties of the (Mn,Ga)As:GaAs granular system and the detailed mechanisms of the formation of (Mn,Ga)As nanocrystals. Even basic characteristics of the (Mn,Ga)As inclusions, such as their compositions, are not yet sufficiently well understood.[12-17] Here we use the notation: (Mn,Ga)As, since both Ga and Mn atoms can occur in the volume of an individual inclusion. Moreover, correct estimation of the exact composition of individual nanocrystals is difficult, due to the small nanocrystal size and the presence of surrounding GaAs matrix. It is likely that the nanoinclusions consist of MnAs with some admixture of Ga, however their magnetic properties, like: Curie temperatures in case of nanocrystal with ferromagnetic properties; and coercieve field values, are rather close to those of MnAs.

In order to answer some of these questions we have carried out: (i) HT annealing of a set of very diluted (Ga,Mn)As samples, that have starting Mn contents of 0.1 to 2%, and the subsequent investigation of their structural and magnetic properties; (ii) HT annealing of (Ga,Mn)As with a starting Mn content of 0.5% in the TEM column for *in-situ* observation of the phase separation process.

In this paper we concentrate on the first set of experiments. We investigate the structural and magnetic properties of (Ga,Mn)As samples with Mn contents increasing from 0.1% (paramagnetic (Ga,Mn)As) to 2% (ferromagnetic (Ga,Mn)As). To our knowledge, the results of a systematic investigation of samples produced by the HT annealing of (Ga,Mn)As with such a low Mn content have not been reported previously. There is also renewed interest in the (Mn,Ga)As:GaAs composite system due to the recently discovered property of (Mn,Ga)As inclusions embedded in semiconducting GaAs that an electromotive force can be induced by the reorientation of the magnetic moments in the (Mn,Ga)As nanocrystals by an external magnetic field.[6]

The results of our annealing experiments performed inside the TEM have been published elsewhere.[19]

## 2. EXPERIMENTAL

Samples studied here have been deposited in a KRYOVAK MBE system dedicated to (Ga,Mn)As. An epi-ready GaAs substrate was glued using In to the Mo holders, which ensures good thermal contact and lateral temperature uniformity during MBE growth. The substrate temperature was monitored using an infrared pyrometer, was the same for all the samples studied, and was chosen to be ~ 270$^o$C. The initial (Ga,Mn)As layers have been grown with an As$_2$ flux generated by a DCA valve cracker effusion cell, with an As/Ga flux ratio of ~ 2. The Mn contents for compositions of 1% or higher was estimated from the increase in growth rate with respect to that of LT GaAs layers grown prior to (Ga,Mn)As deposition by using the period of RHEED intensity oscillations.[20] For lower Mn compositions this parameter was estimated by extrapolation of the temperature dependence of the Mn flux. The thickness of the (Ga,Mn)As layers was 1 μm for Mn contents of 0.1 and 0.3%; 0.7 μm for 0.5% and 0.4 μm for 1 and 2% Mn. After MBE growth and removal from the vacuum system, the samples were each divided into 4 pieces for annealing at different temperatures, with one piece left in the as-grown state. The samples were annealed at three different temperatures: 400, 560 and 630$^o$C, with the annealing temperature controlled by using an infrared pyrometer. For each temperature the samples were annealed simultaneously on the same Mo-holder, which was mounted in the MBE growth chamber. Annealing at the highest temperature of 630$^o$C was carried out in the presence of an As$_2$ flux, to prevent the surfaces of the annealed samples from degrading due to As desorption.[21] The annealing was 40 min for annealing at 400$^o$C and 1 hour for annealing at 560 and 630$^o$C.

The lattice constant in the growth direction (the perpendicular lattice parameter) was measured using high resolution X-ray diffraction, in a Philips X'pert diffractometer. Structural characterization and chemical analysis were carried out on cross-sectional TEM specimens prepared using conventional mechanical polishing and Ar ion milling. The TEM specimens were finished at low ion energies (<1 keV) in order to minimize ion beam induced sample preparation artifacts. Both image and probe aberration-corrected TEM and scanning TEM (STEM) studies were carried out using FEI Titan microscopes operated at 300 kV.

Temperature-dependent magnetization measurements (shown in Sec. IV below) were performed using a commercial superconducting quantum interference device (SQUID) magnetometer from Quantum Design Inc. Magnetic relaxation studies were carried out using a non-commercial low-field SQUID magnetometer.[22] A small superconducting magnet with a time constant of ~ 1 ms delivered small magnetic fields, and was employed to measure of the magnetization magnitude 0.2-0.3 s after field switching. The initial cooling rate to reach the measurement temperature for the relaxation experiments was about 5 K/min. To improve the thermal contact between the thermometer and the top of surface, rather than the backside of the GaAs substrate, the (Ga,Mn)As layer was glued to a sapphire rod connected to the thermometer using a silver paste.

## 3. STRUCTURAL PROPERTIES

The in-plane lattice parameter of the layers is identical to that of the GaAs(001) substrate, since layers are coherently strained to the GaAs substrate.[23] Figure 1 shows the $2\theta/\omega$ X-ray diffractometer scans for 006 symmetric reflections for samples with 0.1, 03, 05, 1 and 2% Mn. The scans are sensitive to strain perpendicular to the sample surface. The 006 reflection was chosen because the angular position of the 004 reflection from (Ga,Mn)As layers with low Mn contents is too close to the 004 diffraction peak from the GaAs substrate. Figure 1 shows the angular positions of the 006 diffraction peaks of the (Ga,Mn)As layers measured before and after annealing at temperatures (Ta) of 400, 560 and 630°C. The intensities have been normalized to that of the 006 reflection of the GaAs substrate (the highest intensity peak at an angular position 2θ of ~109.65°).

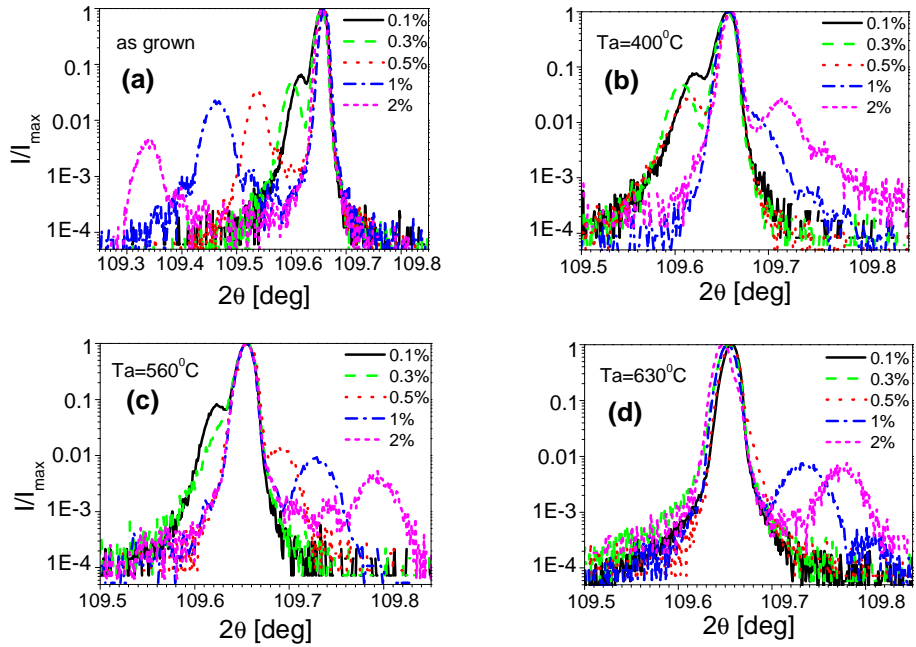

**Fig. 1** (color online) 006 reflections in 2θ/ω X-ray diffractometer scans, measured for (a) as-grown, and (b), (c), (d) 400, 560 and 630°C annealed (Ga,Mn)As layers containing 0.1, 0.3, 0.5, 1 and 2% Mn.

For the as-grown samples, the observed lattice expansion of the (Ga,Mn)As ternary alloy, with respect to GaAs results from Mn partially replacing Ga in the GaAs host lattice.[18,23] However a strong contribution from Mn atoms located at interstitial positions has been also theoretically predicted and then experimentally observed.[23-26] After HT annealing, the difference between the lattice constant of the layer originally constituting the (Ga,Mn)As ternary alloy and the lattice constant of the GaAs substrate is due *only* to strain in the GaAs host lattice caused by the presence of Mn-rich clusters.[27-29] The clusters themselves do not contribute to the diffraction peaks shown in Fig. 1.

It is known from previous reports, that Mn-rich nanocrystals can exert a compressive strain on a surrounding GaAs matrix.[27-29] Since the in-plane lattice parameter of the layer is fixed to that of the GaAs substrate and the nanocrystals cause the local contraction of the surrounding GaAs crystal lattice, the formation of clusters causes a reversal in the sign of the strain from negative (compressive) to positive (tensile strain). This transition can be seen in Fig. 2 which shows the strain calculated from the lattice constants obtained from angular positions of diffraction peaks shown in Fig. 1. The relaxed lattice constant values used for strain evaluation were calculated based on the fact that the in-plane lattice parameters of the layers are identical to that of the GaAs substrate.[23] Strain is defined here as:

$$\varepsilon = (a_s - a_l) / a_l \qquad (1)$$

where:

$a_s$ is the lattice constant of the GaAs substrate

$a_l$ is the relaxed lattice constant of the layer

Values of $a_l$ were calculated from the measured values of $a_\perp$, the lattice parameter of the layer in the direction perpendicular to the substrate, by using the formula:

$$a_l = (a_\perp + a_\parallel \cdot 2b)/(1 + 2b) \qquad (2)$$

where

$a_\parallel = a_s$

$b = C_{12} / C_{11}$

and $C_{11}$ and $C_{12}$ are the elastic constants of the layer material, which are assumed to be the same as for GaAs and take the values:[23] $C_{11} = 11.82 \times 10^{10}$ Pa, $C_{12} = 5.326 \times 10^{10}$ Pa.

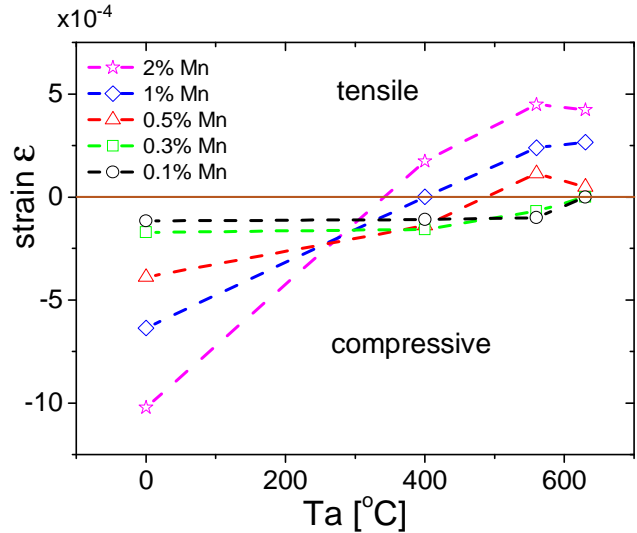

**Fig.2** (color online) Evolution of strain in the as-grown and annealed (Ga,Mn)As layers. Annealing temperatures ($T_a$) were set to 400, 560 and 630°C. The dashed lines joining the experimental points are guides to the eye and do not necessarily follow the true values of strain in layers annealed at intermediate temperatures.

For (Ga,Mn)As layers containing 1 and 2% Mn, the reversal in the sign of the strain already occurs after annealing at the lowest temperature (400°C). Thus this relatively low annealing temperature is already thought to result in phase separation and the formation of (Mn,Ga)As clusters. For (Ga,Mn)As with a Mn content of 0.5% strain reversal occurs at a higher annealing temperature (560°C), and was not observed for samples with 0.1 and 0.3% Mn, presumably due to the much lower internal stress associated with the lower density of clusters. The magnitude of the strain for 1 and 2% Mn is reduced by about 50% upon HT annealing (see Fig. 2). For 0.1 and 0.3% of Mn, the angular positions of the diffraction peaks from the tensile strained layers, after annealing at 630°C, may be too close to the position of the 006 peak of the GaAs substrate to be distinguishable. The apparent tendency for the temperature for strain reversal to increase with decreasing Mn content in HT-annealed (Ga,Mn)As, which can be seen in Fig. 2, may be caused by the higher thermal stability of (Ga,Mn)As with a lower Mn content, or by the contribution of As antisite defects to the (Ga,Mn)As lattice constant, noticeable for Mn contents lower than ~1%.[25]

It was demonstrated previously,[29,30] that strain in the GaAs matrix surrounding (Mn,Ga)As inclusions is larger for zinc-blende clusters than for hexagonal ones. Since the angular positions of the 006 diffraction peaks after annealing at 560°C, are higher than angular positions of the same peaks after annealing at 630°C (see Fig.1c,d), we can infer that the 560°C annealed samples contain a greater proportion of cubic Mn-rich clusters than the samples annealed at 630°C.

More detailed information concerning the local structure of three of the samples containing 0.1, 0.5 and 2 % Mn and annealed at 400 and 560, 630°C have been obtained from TEM investigations. Structural parameters of the annealed samples are summarized in Table 1.

|  | *average size* | | *density [x10$^{-6}$ nm$^3$]* | | *voids* | | *As* | | *cubic (Mn,Ga)As* | | *hexagonal (Mn,Ga)As* | |
|---|---|---|---|---|---|---|---|---|---|---|---|---|
| *Ta* %Mn | 560°C | 630°C | 560°C | 630°C | 560°C | 630°C | 560°C | 630°C | 560°C | 630°C | 560°C | 630°C |
| 0.5 % | 9.8 | 10.8 | 8 | 23 | Yes | Yes | Yes | No | Yes | Yes | Yes | Yes |
| 2 % | 3.9 | 8.8 | 44 | 27 | Yes | Yes | No | No | Yes | Yes | Yes | Yes |

**Tab.1.** Parameters of individual (Mn,Ga)As nanocrystals in HT annealed (Ga,Mn)As layers with Mn content of 2% and 0.5%, estimated from transmission electron microscopy studies.

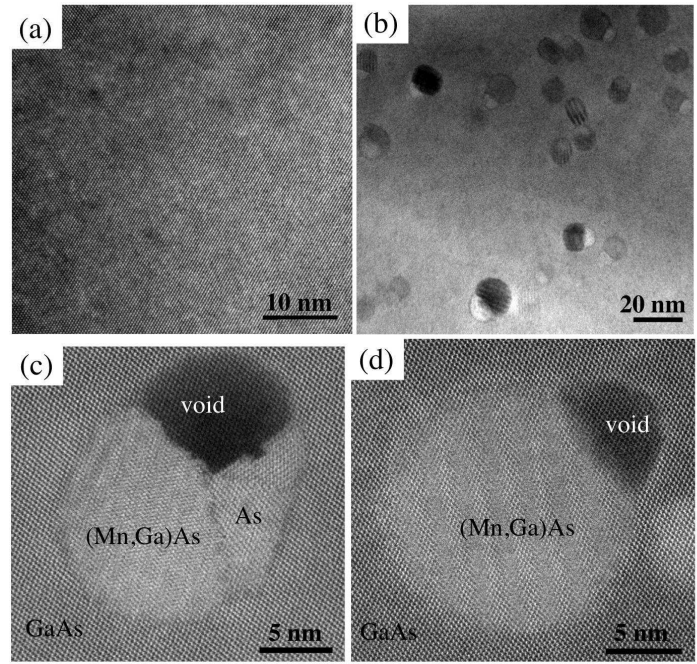

**Fig. 3.** (a, b) - cross-sectional bright-field TEM images of $Ga_{0.995}Mn_{0.005}As$ layer annealed at 400°C and 630°C, respectively. (c) - aberration-corrected ADF STEM image of a void, hexagonal (Mn,Ga)As and rhombohedral As nanocrystals in sample annealed at 560°C. (d) - aberration-corrected ADF STEM image of a void and cubic (Mn,Ga)As nanocrystal in GaAs host. The ADF inner detector semi-angle used was 47.4 mrad.

Figure 3 shows representative TEM images of (Ga,Mn)As layers with an initial Mn content of 0.5 % that had been annealed at different temperatures. In the TEM image shown in Fig. 3(a), the structure of the layer annealed at 400 °C appears to be homogenous. However, nm-sized regions with dark contrast are present, which may indicate the start of a phase separation process. The dark areas may also have resulted from the cross-sectional sample preparation process, even though the specimen was finished at low ion energy. Figure 3(b) shows that high temperature annealing induces nanocrystal formation. The image was recorded slightly underfocus in order to enhance the contrast of the edges of the particles. Interestingly, regions of bright contrast are visible adjacent to many of the nanocrystals. Our detailed electron microscopy analysis[19] reveals that these features are voids. Similar regions can be observed in BF images of low-temperature grown and high-temperature annealed Mn-doped GaAs layers recorded by other groups[12,17,28], however the voids have not been addressed. The structures of the nanocrystals were determined using nano-beam electron diffraction to be as cubic (zinc-blende, ZnS-type) and hexagonal (NiAs-type)[19] with two types of crystal structure coexisting after annealing for 1 hour both at 560 and 630°C. This situation is different from that typically observed in HT-annealed (Ga,Mn)As with a higher Mn content (5% and above), in which zinc-blende clusters are usually identified after annealing at lower temperatures (550°C and below) and hexagonal clusters are observed after annealing at higher temperatures (600°C and above).

In addition, we identified identify rhombohedral (space group 166, symbol *R-3m*) and orthorhombic (space group 64, symbol *Bmab*) As nanocrystals in annealed samples that had been doped with less than 1% of Mn. Figure 3(c) shows an aberration-corrected high-resolution annular dark-field (ADF) scanning TEM (STEM) image of a void, hexagonal (Mn,Ga)As, and rhombohedral As nanocrystals embedded in GaAs in $Ga_{0.995}Mn_{0.005}As$ layer that have been annealed at 560°C. The nanocrystal is associated with void that exhibited dark contrast in ADF STEM image. The As phases were identified using a combination of TEM, STEM images and energy dispersive X-ray spectroscopy signals. Fewer than 5 % of the precipitate complexes contain such As nanocrystals. Formation of As precipitate in low-temperature grown GaAs layers during high-temperature annealing is well known and has been reported for ex. in[31] and in the references therein. Our TEM results suggest that As nanocrystals could form in a low-doped (< 1 % of Mn) (Ga,Mn)As layers as a function of the annealing temperature and Mn concentration. Interestingly As nanocrystals were found only adjacent to hexagonal (Mn,Ga)As nanocrystals and voids. Figure 3(d) show a void and cubic (Mn,Ga)As nanocrystal in $Ga_{0.995}Mn_{0.005}As$ layer that have been annealed at 630°C. Vertical

Moiré fringes are visible at the position of the nanocrystal due to the different lattice parameters of the cubic (Mn,Ga)As than that of GaAs. Interestingly, despite the lattice mismatch expected between the cubic nanocrystal and the GaAs host, no misfit dislocation formation was observed, as shown in Fig. 3 (d).

The annealing of $Ga_{0.98}Mn_{0.02}As$ sample at 560 and 630°C induces the formation of hexagonal and cubic (Mn,Ga)As nanocrystals, as shown in Fig. 4. At 560°C, the average size of the crystals is lower than in $Ga_{0.995}Mn_{0.005}As$ annealed at same temperature, as presented in Table 1. Arsenic precipitates were not observed in this sample. In the case of the hexagonal (Mn,Ga)As nanocrystals, the c-axis is parallel to one of the {111} planes of GaAs, while the cubic nanocrystals are always coherent with the surrounding GaAs matrix.

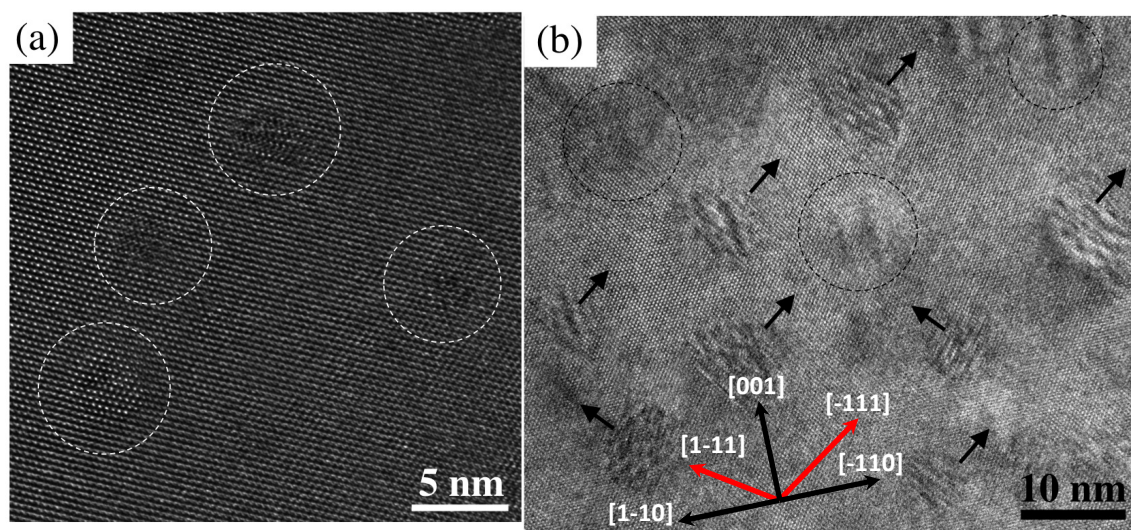

**Fig. 4**. High-resolution aberration-corrected TEM images of hexagonal and cubic (Mn,Ga)As nanocrystals in annealed $Ga_{0.98}Mn_{0.02}As$ layer at (a) 560°C and (b) 630°C. Dashed circles mark the location of the nanocrystals in (a). Dashed circles in (b) mark the cubic (Mn,Ga)As nanocrystals, while the arrows indicate the c-axis of hexagonal crystals that is parallel with one of the {111} orientation of GaAs host.

## 4. MAGNETIC PROPERTIES AND DISCUSSION

Figure 5 shows the temperature dependence of magnetization for the different (Ga,Mn)As samples, recorded for (a) the as-grown and (b,c,d) the annealed layers. The temperature dependence of the magnetization of the as-grown samples with Mn contents of 0.5% and higher, is typical for that expected for thick (Ga,Mn)As layers, with a clear onset of ferromagnetism.[32] No significant magnetic signal is observed for the as-grown layers with 0.1 and 0.3% Mn. A weak magnetic signal of the order of 0.1 kA/m or below was observed for all of the layers after annealing at 400°C, while some of the pieces that have been annealed at higher temperatures exhibit larger magnetic signals.

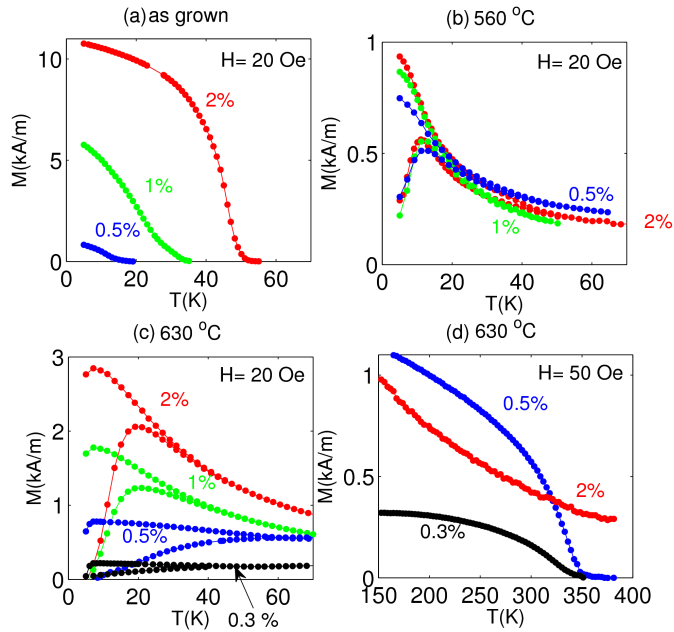

**Fig. 5.** (color online) Temperature $T$ dependence of the (a,d) field-cooled (FC) and (b,c) zero-field cooled/field-cooled (ZFC/FC) magnetization $M$ of all of the layers measured (a) before and (b-d) after heat treatment. (d) temperature dependence of the field-cooled magnetization at higher temperatures for the layers annealed at 630°C. No significant magnetic signal is observed for the as-grown or annealed layers with 0.1% Mn. In both the as-grown and the annealed cases, $M$ is converted to units of kA/m by considering the initial volumes of the (Ga,Mn)As layers.

In Fig. 5(b), the layers with the highest Mn contents have similar magnetic responses, with a maximum in zero-field cooled (ZFC) magnetization around $T = 15$ K, and a paramagnetic-like field-cooled (FC) magnetization, suggesting a superparamagnetic behavior. Similar curves are observed in Fig. 5(c) for the 1 and 2% of Mn samples annealed at the highest temperature (630°C), albeit with a maximum near 20 K. If we consider this temperature as a blocking temperature $T_b$, we can estimate the anisotropy constant of the particles: Assuming that the Arrhenius law: $\ln(\tau/\tau_0) = KV/k_B T$ is followed with $\tau$ of about 30 s (that is the time scale of a typical temperature-dependent magnetization measurement) and $\tau_0$ about $10^{-10}$ s (that is the typical value for nanoparticles of several nanometers, see e.g. Ref. 33), we obtain that $KV/k_B T_b = 26.43$. With $T_b = 20$ K, and a diameter of 10 nm for the particles, we obtain $K = 13930$ J/m$^3$ (or 139300 erg/cm$^3$), in agreement with earlier determinations.[34]

The low-temperature FC magnetization of both layers is flatter than that measured for 560°C annealed samples, suggesting increased magnetic interaction and possible spin-glass behavior.[33] The FC magnetization curve of the layer with 0.5% Mn is even flatter. The layer with 0.3% Mn, which had not displayed any sizable magnetism in the as-grown state and after annealing at 560°C, is found to exhibit a magnetic response similar to that of the layer with 0.5% Mn after annealing at 630°C. A step decrease in magnetization at low temperature, (below 6-7 K) can be observed in these measurements for the layers annealed at the highest temperature. Rather than the effect of magnetic (dipolar) interaction, we believe that the reduction of the magnetization is associated with the annealing-induced diffusion of the In that was used to fix the GaAs substrate to its holder in the MBE chamber. We have performed magnetization measurements from lower temperatures (2 K, not shown) which show that the magnetization below 6-7K decreases and becomes negative, still decreasing until it becomes temperature independent. This behavior is typical for diamagnets and we believe that it reflects the superconducting behavior associated with the diffusion of In as discussed in Ref. 35.

Interestingly, in Fig. 5(d), which shows FC magnetization at higher temperatures, the layers that showed superparamagnetic-like behavior (1 and 2% Mn) do not show any further features at higher temperatures, while those that showed "flat FC magnetization" (0.3 and 0.5% Mn) appear to undergo magnetic phase transitions near $T = 350$ K. Similar high-temperature ferromagnetic phase transitions have previously been observed in HT annealed (Ga,Mn)As systems and were then associated with Mn-rich nanocrystals with cubic (zinc-blende) structures, with have a higher $T_C$ than the MnAs in its natural hexagonal phase.[30] Our results suggest that HT annealing yields different microscopic configurations and crystal structures of (Mn,Ga)As inclusions for layers with more than 1% Mn than in layers with less than 1% Mn. It is interesting to observe that, after annealing at 560°C the samples with 0.5, 1 and 2% Mn exhibit almost the same $M(T)$ dependencies, *i.e.* superparamagnetic behavior with a similar blocking temperature close to 15 K. We assume that the formation of (Mn,Ga)As precipitates during HT annealing starts with the nucleation of small cubic clusters. Upon increasing annealing temperature, the small cubic clusters coalesce into larger ones. Then, depending on the Mn content either size increases further, or they undergo transitions from the cubic to the hexagonal phase. Since bulk, pure MnAs does not exist in a zinc-blende structure, above a certain critical size the clusters adopt only the hexagonal phase, providing that they are sufficiently Mn-rich. This critical size is close to 15 nm (see Fig. 3). Zinc-blende (Mn,Ga)As nanocrystals with larger sizes were not observed either in our TEM images, or in TEM results published by other groups.[9,27,28,30]

The magnetic properties of granular system with two kinds of nanoparticles are not trivial to understand. In ferrofluids composed of single-domain magnetic nanoparticles separated by surfactants, superparamagnetism is usually observed in systems in which the nanoparticles do not interact magnetically.[33] If the nanoparticles interact because of their sizes or spacings, then they may instead display spin-glass behavior.[33] In the present study the most concentrated systems (1 and 2% Mn) show superparamagnetic behavior up to high temperatures, while the layers that contain less Mn appear to exhibit high-temperature magnetic transitions.

The dynamic (time-dependent) magnetic properties of layers with 2% Mn that had been annealed at higher temperature were investigated in more detail. The results of these measurements are shown in Figs. 6 and 7.

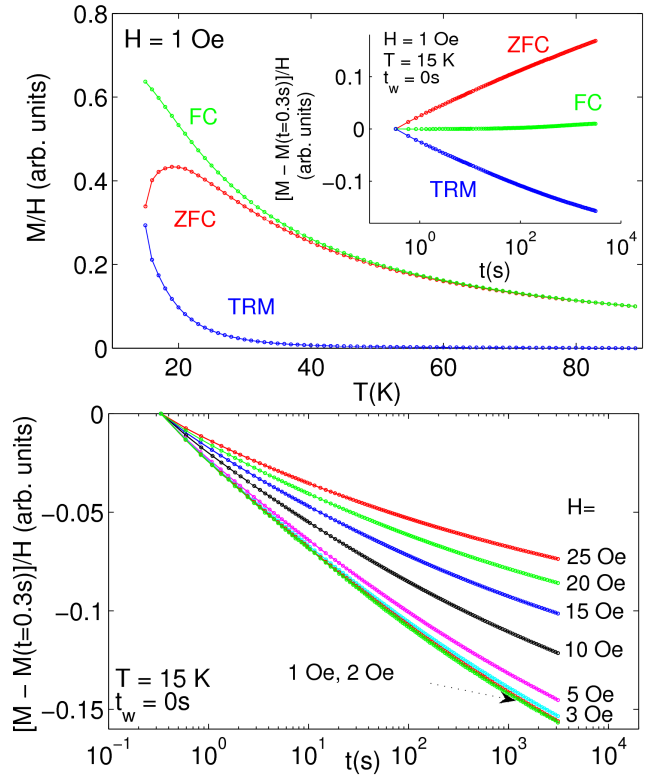

**Fig. 6.** (color online) Upper panel: Temperature (main frame) and time (inset) dependence of the ZFC, FC and TRM magnetization of the layer with 2% Mn, plotted as *M/H* and [*M* − *M*(*t* = 0.3 s)]/*H*, respectively. Lower panel: TRM relaxation recorded at $T_m$ = 15 K for different magnetic fields.

In addition to usual temperature-dependent zero-field cooled (ZFC), field cooled (FC), and thermoremanent magnetization (TRM) measurements, the top panel in Fig. 6 also shows time-dependent ZFC, FC, and TRM relaxation measurements performed at constant temperature. If small enough magnetic fields are used, then the response of the system is linear and the system is only probed by the magnetic field. The lower panel in Fig. 6, suggests that a linear response is achieved only when the magnetic field is lower than $H \sim 3$ Oe. At lower fields, $M/H$ is approximately independent of the magnetic field. We therefore chose $H = 1$ Oe to perform our experiments. The inset to the top panel of Fig. 6 shows that the relaxation curves essentially obey the superposition relation: $M_{ZFC} \sim M_{FC} - M_{TRM}$.[36] We therefore considered the relaxation of the TRM magnetization instead of the ZFC magnetization, as is more usual in studies of *e.g.*, spin-glass systems, in order to record the magnetization in zero magnetic field and thus limit the contribution from the diamagnetic substrate.

Figure 7 shows relaxation curves of TRM, recorded after rapidly cooling the layer with 2% Mn to low temperature in a small magnetic field and recording the evolution of the magnetization with time in zero magnetic field while keeping the temperature constant. For spin glasses, such relaxation curves depend on the history of the system, and on how long one waits before switching off the magnetic field and recording the magnitude of the magnetization.[37] During such waiting time, the magnetic configuration of the spin glass is rearranged towards its equilibrium configuration without ever reaching it; the spin glasses ages. Such an aging phenomenon is not observed in the case of a superparamagnet, whose magnetic relaxation is related mainly to thermally activated processes associated with the magnetic anisotropies of the individual particles.[33]

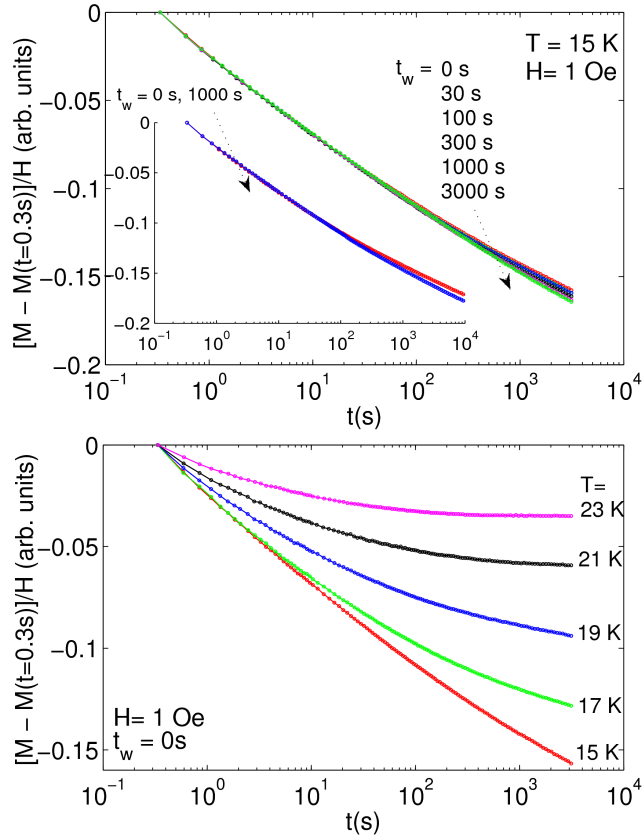

**Fig. 7.** (color online) Time *t* dependence of the thermoremanent magnetization (TRM) of the layer with 2% Mn, plotted as $[M - M(t = 0.3\text{ s})]/H$. The sample was cooled rapidly from a reference temperature $T_{\text{ref}} = 120$ K to the measurement temperature $T_m$ in a magnetic field $H = 1$ Oe. After a waiting time $t_w$, the magnetic field was switched off and the magnetization was recorded as a function of time while keeping the temperature constant. In the top panel, the TRM magnetization was recorded at $T_m = 15$ K for different waiting times (the inset shows examples of measurements recorded over longer time scales). In the lower panel, the TRM magnetization was recorded at different temperatures without a waiting time, *i.e.*, for $t_w = 0$ s.

The top panel in Fig. 7 shows that TRM relaxation curves recorded at low temperature are essentially waiting-time independent, confirming superparamagnetic behavior. Spin-glass memory experiments[37] were also performed and did not reveal aging, memory or rejuvenation effects that would be typical for spin-glasses. The lower panel in Fig. 7 shows that superparamagnetic relaxation is maintained up to a temperature just above that at which a cusp is observed in the ZFC magnetization.

## 5. CONCLUSIONS

MBE grown layers of a (Ga,Mn)As ternary alloy with Mn contents, of 0.1, 0.3, 0.5, 1 and 2% have been subjected to high temperature post-growth annealing at temperatures of 400, 560 and 630$^o$C. Annealing of the layers with these low Mn contents leads to the formation of a phase separated nanocomposite system in which Mn-rich nanocrystals are buried in the GaAs matrix, with a much lower density than that reported previously. X-ray diffraction, TEM and SQUID magnetometry are used to show that the phase separation process is already initiated at a temperature as low as 400$^o$C, by the formation of nanoscale (nm sized) Mn-rich nanocrystals which have a zinc-blende structure and coalesce into larger (5-15 nm) crystals at higher annealing temperatures. In high temperature annealed (Ga,Mn)As with a low Mn contents (0.5%) both cubic (zinc-blende) and hexagonal (NiAs-type) crystals coexist, even after annealing at temperatures as high as 630$^o$C. The maximum size of the cubic nanocrystals is limited to about 15 nm. Moreover, the annealing of low Mn content (Ga,Mn)As layers can result in a more complex structure consisting of hexagonal (Mn,Ga)As, As nanocrystals, and voids. The minimum Mn concentration that results in detectable phase separation and in ferromagnetic or superparamagnetic properties in the HT annealed (Ga,Mn)As is close to 0.3%. For the samples with Mn contents of below 1% that contain both cubic and hexagonal crystals besides the signature of a superparamagnetic phase, a ferromagnetic phase transition with a $T_C$ of about 350 K is observed. In high temperature annealed (Ga,Mn)As with a higher Mn contents (1 and 2%), only superparamagnetic properties are observed. Measurements of magnetization relaxation over the time spans of 0.1 – 10 000 s corroborate the observation of superparamagnetic behavior of the (Mn,Ga)As nanocrystals, as well as the absence of spin-glass dynamics.

## ACKNOWLEDGMENTS

This work was partly supported by the "FunDMS" Advanced Grant of the European Research Council within the "Ideas" 7$^{th}$ Framework Programme of the European Commission. The Swedish Research Council (VR) and the Göran Gustafsson Foundation (Sweden) are also acknowledged for financial support of the MBE system at MAX-Lab, Lund and the SQUID magnetometers at Uppsala University.

[1] H. Ohno, Nature Mater. **9**, 952 (2010).

[2] T. Dietl, Nature Mater. **9**, 965 (2010).

[3] A. Bonanni and T. Dietl, Chem. Soc. Rev. **39**, 528 (2010).

[4] H. Katayama–Yoshida, K. Sato, T. Fukushima, M. Toyoda, H. Kizaki, V. A. Dinh, and P. H. Dederichs, Phys. Status Solidi (a) **204**, 15 (2007).

[5] T. Dietl, J. Appl. Phys. **103**, 07D111 (2008).

[6] D. C. Ralph, Nature 474, E6 (2011).

[7] S. Kuroda, N. Nishizawa, K. Takita , M. Mitome, Y. Bando, K. Osuch, and T. Dietl, Nature Mater. **6**, 440 (2007).

[8] A. Bonanni, A. Navarro-Quezada, Tian Li, M. Wegscheider, Z. Matej, V. Holy, R.T. Lechner, G. Bauer, M. Rovezzi, F. D'Acapito, M. Kiecana, M. Sawicki, and T. Dietl, Phys. Rev. Lett. **101**, 135502 (2008).

[9] M. Kaminska. A. Twardowski, and D. Wasik, J. Mater Sci: Mater Electron **19**, 828 (2008).

[10] R. P. Campion, K. W. Edmonds, L. X. Zhao, K. Y. Wang, C. T. Foxon, B. L. Gallagher, and C. R. Staddon. J. Crystal Growth **247**, 42 (2003).

[11] D. Chiba, Y. Nishitani, F. Matsukura, and H. Ohno, Appl. Phys. Lett. **90**, 122503 (2007).

[12] J. De Boeck, R. Oesterholt, A. Van Esch, H. Bender, C. Bruynseraede, C. Van Hoof, and G. Borghs, Appl. Phys. Lett. **68**, 2744 (1996).

[13] H. Akinaga, S. Miyanishi, K. Tanaka, W. Van Roy, and K. Onodera, Appl. Phys. Lett. **76**, 97 (2000).

[14] M. Moreno, B. Jenichen, L. Däweritz, and K. H. Ploog, J. Vac. Sci. Technol. B **23**, 1700 (2005).

[15] A. Kwiatkowski, D. Wasik, M. Kamińska, R. Bożek, J. Szczytko, A. Twardowski, J. Borysiuk, J. Sadowski, and J. Gosk, J. Appl. Phys. **101**, 113912 (2007).

[16] K. Lawniczak-Jablonska, J. Libera, A. Wolska, M. T. Klepka, P. Dłużewski, J. Sadowski, D. Wasik, A. Twardowski, A. Kwiatkowski, and K. Sato, Phys. Status Solidi RRL **5**, 62 (2011).

[17] K. Lawniczak-Jablonska, J. Bak-Misiuk, E. Dynowska, P. Romanowski, J.Z. Domagala, J. Libera, A. Wolska, M.T. Klepka, P. Dluzewski, J. Sadowski, A. Barcz, D. Wasik, A. Twardowski, A. Kwiatkowski, J. Solid State Chemistry **184,** 153 (2011).

[18] H. Ohno, A. Shen, F. Matsukura, A. Oiwa, A. Endo, S. Katsumoto, and Y. Iye, Appl. Phys. Lett. **69**, 363 (1996).

[19] A. Kovács, J. Sadowski, T. Kasama, J. Z. Domagala, R. Mathieu, T. Dietl, and R. E. Dunin-Borkowski, J. Appl. Phys. **109**, 083546 (2011).

[20] J. Sadowski, J. Z. Domagala, J. Bak-Misiuk, S. Kolesnik, M. Sawicki, K. Swiatek, J. Kanski, L. Ilver, and V. Ström, J. Vac. Sci.Technol. B **18**, 1697 (2000).

[21] Z. Y. Zhou, C. X. Zheng, W. X. Tang, D. E. Jesson, and J. Tersoff, Appl. Phys. Lett. **97**, 121912 (2010).

[22] J. Magnusson, C. Djurberg, P. Granberg, and P. Nordblad, Rev. Sci. Instrum. **68**, 3761 (1997).

[23] I. Kuryliszyn-Kudelska, J. Z. Domagala, T. Wojtowicz, X. Liu, E. Łusakowska, W. Dobrowolski, and J. K. Furdyna, J. Appl. Phys. **95**, 603 (2004).

[24] J. Masek, J. Kudrnovsky, and F. Maca, Phys. Rev. B **67**, 153203 (2003).

[25] J. Sadowski, and J. Z. Domagala, Phys. Rev. B **69**, 075206 (2004).

[26] L. X. Zhao1, R. P. Campion, P. F. Fewster, R. W. Martin, B. Ya Ber, A. P. Kovarsky, C. R. Staddon, K. Y. Wang, K. W. Edmonds, C. T. Foxon, and B. L. Gallagher, Semicond. Sci. Technol. **20**, 369 (2005).

[27] M. Moreno, B. Jenichen, L. Däweritz, and K. H. Ploog, J. Vac. Sci. Technol. B **23**, 1700 (2005).

[28] M. Moreno, V. M. Kaganer, B. Jenichen, A. Trampert, L. Däweritz, and K. H. Ploog, Phys. Rev. B **72**, 115206 (2005).

[29] J. Bak-Misiuk, J.Z. Domagala, P. Romanowski, E. Dynowska, E. Lusakowska, A. Misiuk, W. Paszkowicz, J. Sadowski, A. Barcz, and W. Caliebe, Radiation Physics Chemistry **78**, S116 (2009).

[30] M. Yokoyama, H. Yamaguchi, T. Ogawa, and M. Tanaka, J. Appl. Phys. **97**, 10D317 (2005).

[31] M. Toufella, P. Puech, R. Carles, E. Bedel, C. Fontaine, A. Claverie, and G. Benassayag,

J. Appl. Phys. 85, 2929 (1999).

[32] J. Sadowski, R. Mathieu, P. Svedlindh, J. Z. Domagala, J. Bąk – Misiuk, K. Swiątek, M. Karlsteen, J. Kanski, L. Ilver, H. Åsklund, and U. Södervall, Appl. Phys. Lett. **78**, 3271 (2001).

[33] P. E. Jönsson, Adv. Chem. Phys. **128**, 191 (2004).

[34] R. S. DiPietro, H. G. Johnson, S. P. Bennett, T. J. Nummy, L. H. Lewis, and D. Heiman, Appl. Phys. Lett. **96**, 222506 (2010).

[35] Y. K. Li, Y. Huang, Z. Fan, C. Jiang, X. B. Mei, B. Yin, J. M. Zhou, J. C. Mao, J. S. Fu, and E. Wu, J. Appl. Phys. **71**, 2018 (1992).

[36] P. Nordblad, L. Lundgren, and L. Sandlund, J. Magn. Magn. Mater. **54-57**, 185 (1986).

[37] P. E. Jönsson, R. Mathieu, P. Nordblad, H. Yoshino, H. Aruga Katori, and A. Ito, Phys. Rev. B **70**, 174402 (2004).